\documentstyle[12pt,epsfig]{article}
     \def\lsim{\raise0.3ex\hbox{$<$\kern-0.75em\raise-1.1ex\hbox{$\sim$}}}
\def\gsim{\raise0.3ex\hbox{$>$\kern-0.75em\raise-1.1ex\hbox{$\sim$}}}
\def\noi{\noindent}

\def\bea{\begin{eqnarray}}  \def\eea{\end{eqnarray}}
\def\beq{\begin{equation}}   \def\eeq{\end{equation}}

\def\beeq{\begin{eqnarray}} \def\eeeq{\end{eqnarray}}

\begin{document}
\begin{center}
\vbox to 1 truecm {}
{\Large \bf }\par 
\vskip 3 truemm
  {\Large \bf Elliptic Flow and Fixed p$_{\bf T}$ Suppression in}\par \vskip 3 truemm
  {\Large \bf a Final State Interaction Model} \\
\vskip 1.5 truecm
{\bf A. Capella}\\ 
Laboratoire de Physique Th\'eorique\footnote{Unit\'e Mixte de
Recherche UMR n$^{\circ}$ 8627 - CNRS}
\\ Universit\'e de Paris XI, B\^atiment 210, 91405 Orsay Cedex, France
\vskip 5 truemm
{\bf E. G. Ferreiro}\\
Departamento de F{\'\i}sica de Part{\'\i}culas, 
Universidad de Santiago de Compostela, 15782 Santiago de Compostela, 
Spain

\end{center}
\vskip 1 truecm
\begin{abstract}
It has been shown that a final state interaction model, used to
describe $J/\psi$ suppression, can also describe the fixed $p_T$
suppression of the $\pi^0$ (and charged pions) yield at all values of $p_T$, with a final
state interaction cross-section $\sigma$ close to one milibarn. We
propose an extension of the model to the pion motion in the
transverse plane -- which introduces a dependence of the suppression on
the azimuthal angle $\theta_R$. Using the same value of $\sigma$, we
obtain values of the elliptic flow $v_{2}$ close to the experimental
ones, for all values of $p_T$, including the soft $p_T$ region.
\end{abstract}

\vskip 2 truecm
\noi LPT-Orsay-06-12\par
\noi February 2006
\newpage
\pagestyle{plain}

\section{Introduction}
\hspace*{\parindent} Statistical QCD (i.e. QCD applied to a system in thermal equilibrium)
on a lattice predicts the existence of a new phase of matter, quark
matter or quark gluon plasma. It is therefore most important to find
signals of the thermalization in the heavy ion data. To achieve
thermalization is by no means trivial. High densities are only a
necessary condition. Strong final state interaction is also needed, as
well as an interaction time long enough for the system to equilibrate.
\par

Nuclear collisions are, a priori, not favorable to produce a
thermalized system. 
%
Indeed, $pA$ collisions
can be described with independent string models \cite{1r}. There is no
indication in the data of final state interaction between particles
produced in different strings -- despite the fact that the proton
collides with several nucleons of the nucleus and densities several
times larger than the proton proton one are reached and that several
strings are produced in a transverse area of 1~fm$^2$.\footnote{This
is probably due to the small transverse size of the string as
characterized by the correlation length of the non-perturbative
interaction (about 0.2~fm from lattice data).} In a central $AA$
collision this string density is higher and ``cross-talk'' between
different strings is expected. Correspondingly, final state interaction is seen in
the data~: strangeness enhancement, $J/\psi$ suppression, large $p_T$
suppression, elliptic flow, etc. cannot be described with independent
strings. It could happen that string interaction is so strong that the very concept of string becomes meaningless. On the contrary, in view of the nuclear transparency observed in $pA$, one could expect that final
state interaction will take place with a comparatively small cross-section.
In this case the bulk of the produced particles is hardly affected by
the final state interaction. Only rare events, of the type mentioned
above, are strongly modified. \par

Heavy ion data lend support to this scenario. For instance, one could
expect that thermalization strongly affects the $p_T$ distribution of
produced particles -- especially at low $p_T$. Experimentally, the
variation of $<p_T>$ between the most peripheral and the most central
collisions is quite small (about 20 \% at RHIC) and consistent with the
small increase observed in $pA$. Also the RHIC data on
$\overline{p}/h^-$ are flat from peripheral to central collisions --
showing no sign of strong $\overline{p}p$ annihilation which would
produce a change in the $\overline{p}/h^-$ ratio for (out of equilibrium) peripheral
collisions followed by a flatness when thermalization is reached.
Turning to rearer events, the centrality dependence of the ratios
$B/h^-$ and $\overline{B}/h^-$ at RHIC shows no sign of saturation --
increasing monotonically from peripheral to central collisions for
$\Lambda$, $\Xi$ and $\Omega$ -- contrary to statistical model
predictions \cite{2r}. Also the decrease of $<p_T^2>$ of $J/\psi$ at
large centralities, predicted in a deconfining phase transition
scenario, is not seen in the data \cite{3r}.\par

One of the main arguments for (early, transverse) thermalization at
RHIC is the description of elliptic flow data with an ideal (no
viscosity) hydrodynamical model. However, this view has recently been
challenged \cite{4r}. Experimentally \cite{5r}, it turns out that ideal
hydrodynamics describes the $v_{2}$ data only in a small low-$p_T$ interval and
a very small rapidity range. Moreover, it only describes the minimum
bias data and fails to reproduce the centrality dependence. In
particular, it overestimates the data at small impact parameter, where
the densities are higher. A review of hydrodynamic model \cite{6r},
coalescence model \cite{7r} and transport model \cite{8r} calculations
and its comparison with experimental data, as well as a comprehensive
list of references on these models can be found in \cite{5r}.\par

Quantitative support to the idea of a weak final state interaction,
which affects only rare events can be found in refs. \cite{9r} and
\cite{10r}. In \cite{9r} it has been shown that the strangeness
enhancement data \cite{11r} can be described with a final state interaction
cross-section $\sigma = 0.2$~mb. In \cite{10r}, the $J/\psi$
suppression at CERN-SPS \cite{12r} is reproduced with $\sigma = 0.65$~mb. More
recently, it has been shown \cite{13r} that the large $p_T$ suppression
of $\pi^0$'s, observed at RHIC can be reproduced with the same final
state interaction model \cite{9r} \cite{10r}, using an effective
cross-section of the order of 1~mb. In this formalism the ratio
$R_{AA}(b,p_T) = dN^{AA}/dy (b, p_T)/n(b) dN^{pp}/dy$ (where $n(b)$ is
the average number of binary collisions at fixed impact parameter $b$)
can be reproduced at all values of $p_T$ -- including the soft $p_T$
region. Moreover, the large $p_T$ results turn out to be rather insensitive to
the value of the $p_T$ shift of the $\pi^0$ resulting from its
interaction with the hot medium. (In our approach the results for charged pions and $\pi^0$'s are identical).\par

In this paper we propose an extension of the final state interaction
model \cite{9r,10r,13r} in order to take into account the motion of
the pion in the transverse plane. This introduces a dependence of
the pion survival probability on its azimuthal angle $\theta_R$,
which, in turn, gives rise to a (positive) contribution to the elliptic
flow $v_{2}$.\par

The fact that the mechanism responsible for the large $p_T$ suppression
does give a contribution to the elliptic flow at large $p_T$
of the order of the experimental one  is known \cite{14r} \cite{15r}.
Here we show that, in the model \cite{9r,10r,13r}, values of
$v_{2}$ very close to the experimental ones are obtained in the
whole $p_T$ region, including the low $p_T$ one, using the same value of the final state interaction
cross-section (of the order of 1~mb) needed to reproduce the
experimental values of $R_{AA}$.\par

Although this contribution to $v_{2}$ results from an asymmetry in the
azimuthal angle, it can be qualified as non-flow \cite{15r}. Indeed,
the mechanism from which it arises (fixed $p_T$ suppression) is maximal
at zero impact parameter and, moreover, thermalization is not needed.

\section{The final state interaction model}
\hspace*{\parindent} As in refs. \cite{9r,10r,13r} the model we are going to use is based
on the gain and loss differential equations which govern final state
interactions. Following \cite{16r} we assume boost invariance and
dilution in time $\tau$ of the (transverse) densities $\rho_i$ due only
to longitudinal expansion ($\rho \sim 1/\tau$). We then have \cite{16r}
\beq
\label{1e}
\tau\ {d\rho_i \over d\tau} = \sum_{k,\ell} \sigma_{k\ell} \ \rho_k\ \rho_{\ell} - \sum_k \sigma_{ik} \ \rho_i\ \rho_k \ .
\eeq

\noi Here $\rho_i \equiv dN^{AA\to i}(b)/dy d^2s$ are transverse
densities and $\sigma_{ij}$ are the final state interaction
cross-sections averaged over the relative velocities of the incoming particles.
The first term of (\ref{1e}) describes the gain in type $i$ particle
yield resulting from the interaction of $k$ and $\ell$. The second one
corresponds to the loss of type $i$ particles resulting from its
interaction with particle $k$. Eqs. (\ref{1e}) have to be integrated
between initial time $\tau_0$ and freeze-out time $\tau_f$. The
solution depends only on the ratio $\tau_f/\tau_0$. Actually particles
$i,j, k, \ell \cdots$ can be either hadrons or partons. Indeed, at early
times the densities are very high and hadrons not yet formed. Thus, at
early times eqs. (\ref{1e}) describe final state interactions at a
partonic level. Only at later times, we have interactions of full
fledged hadrons and, thus,  $\sigma$ represents an effective cross-section
averaged over the interaction time. For this reason 
formation times -- which are not introduced here -- do not play 
an important role.\par

Let us now consider a $\pi^0$ produced at fixed $p_T$ interacting with
the hot medium consisting of charged plus neutrals. We assume that in
the interaction, with cross-section $\sigma$, the $\pi^0$ suffers a decrease in its transverse momentum with a
$p_T$-shift $\delta p_T$. This produces a loss in the $\pi^0$ yield
in a given $p_T$ bin. There is also a gain resulting from $\pi^0$'s
produced at $p_T + \delta p_T$. Due to the steep fall-off of the $p_T$
spectrum the loss is larger than the gain, resulting in a net
suppression of the $\pi^0$ yield at a given $p_T$. In this case there
is a single final state interaction channel and eqs. (\ref{1e}) can be
solved analytically. Integrating both members of eq. (\ref{1e}) over
$d^2s$, we obtain the following expression for the survivable
probability of the $\pi^0$ in the medium
\beq
\label{2e}
S_{\pi^0}(y, p_T, b) = \exp \left \{ - \sigma \ \rho (b,y) \left [ 1 - {N_{\pi^0} (p_T + \delta p_T)\over N_{\pi^0}(p_T)} (b)\right ] \ \ell n (\rho (b,y)/\rho_{pp}(y))\right \} \ .
\eeq

\noi  Here $N_{\pi^0}(p_T)$ denotes the yield of $\pi^0$'s at a given
$p_T$ (see section 3). With $\delta p_T \to \infty$, the quantity
inside the square bracket is equal to one and the gain term vanishes.
In this case, the survival probability has the same expression as in
the case of $J/\psi$ suppression without $c\overline{c}$ recombination.
With $\delta p_T = 0$, the loss and gain terms are identical and the
survival probability is equal to one.\par

The density $\rho (b, y)$ in eq. (\ref{2e}) is given by 
\beq
\label{3e}
\rho (b, y) = {dN^{AA} \over dy} (b)/G(b)
\eeq

\noi where the inclusive $AA$ distribution refers to the production of
charged plus neutrals and $G(b)$ is an effective transverse area (see below). \par

The argument of the logarithm in eq. (\ref{2e}) is the ratio of the freeze-out $\tau_f$ to the initial time $\tau_0$. The solution of (\ref{1e}) depends only on the ratio $\tau_f/\tau_0$. We use the inverse
proportionality between time and density (see above) to write
\beq
\label{4e}
\tau_f / \tau_0 = \rho_0/\rho_f = \rho (b, y)/\rho_{pp}(y)
\eeq

\noi where the freeze-out density $\rho_f$ has been taken to be the $pp$ one, with 
\beq
\label{5e}
\left . \rho_{pp}(y) = 3\ {dN^{pp \to h^-} \over dy}\right / \pi R_p^2 \ .
\eeq 

\noi At $\sqrt{s} = 200$~GeV, $\rho_{pp}(y=0) = 2.24$~fm$^{-2}. $\footnote{With this prescription the average value of $\tau_f/\tau_0$ is of $5\div 6$. Thus, if $\tau_0 \sim 1$~fm we obtain a freeze out time $\tau_f \sim$ 5-6~fm.}\par

The densities in eqs. (\ref{1e})-(\ref{4e}) are computed in the dual
parton model \cite{1r}. They are given by a linear combination of the
average number of participants and of binary collisions. The
coefficients are obtained from convolutions of momentum distribution
and fragmentation functions. Their numerical values for $Au$ $Au$ collisions at $\sqrt{s} =
200$~GeV at each $b$ are given in \cite{18r}. Shadowing corrections,
described in detail in \cite{18r}, are included in the calculation. Finally the geometrical factor $G(b)$, resulting from
the integration over $d^2s$ of the two members of eqs. (\ref{1e}), is
given by
\beq
\label{6e}
\left . G^{-1}(b) = \int d^2s {dN^{AA} \over dy \ d^2s} n(b,s) \right /  {dN^{AA} \over dy}(b) \ n(b)\
\eeq

\noi where $n(b,s) = \sigma_{pp}T_A(s)T_A(b-s)/\sigma_{AA}(b)$. The nuclear profiles $T_A$ are computed from Woods-Saxon densities \cite{17r}. We use the non-diffractive inelastic cross-section $\sigma_{pp} = 34$~mb at $\sqrt{s} = 200$~GeV. In writing eq. (\ref{6e}) we have assumed that the large $p_T$ probe scales with the number $n$ of binary collisions\footnote{In the dual parton model at very high energies,
$dN^{AA}/dyd^2s$, before shadowing corrections are introduced, is
also proportional to the number of binary collisions. In this case, $G^{-1}(b) =
\int d^2s [T_{AA}(b,s)]^2/T_{AA}^2(b)$ is a purely geometrical factor. The results obtained with these two definitions of $G(b)$ are quite similar.}.

\section{Numerical results for the $\pi^0$ suppression}
\hspace*{\parindent}The suppression of the $\pi^0$ yield at fixed $p_T$ is usually
presented via the ratio
\beq 
\label{7e}
\left . R_{AA}(b,p_T) = {dN^{AA} \over dy} (b, p_T)\right / n(b) \  {dN^{pp} \over dy} (p_T) \ .
\eeq

\noi In this paper we shall concentrate on
the mid-rapidity region $|y^*| < 0.35$. In order to compute this ratio
we have to know its value $R_{AA}^0(b, p_T)$ in the absence of final
state interaction. Its $p_T$ integrated value (close to 1/3 for the 10 \% most central $Au$ $Au$ collisions) is obtained in the dual parton model with shadowing corrections \cite{13r}.
Its $p_T$ dependence ($p_T$ broadening) is controlled by two
mechanisms. The first one is the decrease of shadowing effects when
$p_T$ increases. This produces an increase of $R_{AA}^0(b, p_T)$
towards unity. The second mechanism is the Cronin effect proper, which produces a rise of $R_{AA}^0(b, p_T)$ above unity. In ref.
\cite{13r} we have used a phenomenological parametrization of both
effects of the form
\beq
\label{8e}
\left . R_{AA}^0(b, p_T) = R_{AA}^0(b, p_T=0) \left ( {p_T + p_0^{AA}(b) \over p_T + p_0^{pp}}\right )^{-n}\right / \left ( {p_0^{AA}(b) \over p_0^{pp}}\right )^{-n}
\eeq

\noi where $p_0(b) = (n-3)/2<p_T>_b$, $n = 9.99$ and $<p_T>_b$ is the
experimental value of $<p_T>$ at each $b$ \cite{19rnew}. Details of the phenomenological parametrization of the $p_T$
distribution leading to eq. (\ref{8e}) can be found in \cite{13r}. The initial suppression
$R_{AA}^0(b, p_T)$ obtained from (\ref{8e}) reaches values substantially larger than unity and does not decrease towards one at large $p_T$ (see Fig.~1~; numerical values are given in \cite{13r}).\par

The value of the $\pi^0$ suppression, characterized by the ratio
(\ref{7e}), 
can be computed by multiplying $R_{AA}^0(b, p_T)$  
by the $\pi^0$ survival probability due to final state
interaction, given by eq. (\ref{2e}):
$R_{AA}(b, p_T)=R_{AA}^0(b, p_T)\ S_{\pi^0}(p_T, b)$. 

In order to compute the survival probability, we have to
define the quantity inside the square brackets in eq. (\ref{2e}). Using the same
parametrization of the $p_T$ distribution as in eq. (\ref{8e}) we have
\beq
\label{9e}
{N_{\pi^0}(p_T + \delta p_T) \over N_{\pi^0Ñ} (p_T)}(b) = \left ( {p_0(b) + p_T + \delta p_T \over p_0 (b) + p_T}\right )^{-n}\ .
\eeq

Note that this quantity represents 
the gain of particles at a given $p_T$ resulting from the particles or partons
produced originally at $p_T+\delta p_T$ and that are shifted due to the interaction with the dense 
medium. As for eq. (\ref{8e}), this parametrization is extracted from the power-law function
proposed in Ref. \cite{19rnew}.

We have studied 
different possibilities for the shift $\delta p_T$ \cite{13r}.
In a way, this is equivalent to a test on how the mean energy loss ($p_T$ being $E$ at $y = 0$) 
behaves with the energy.
We have also tried different parametrizations for $R_{AA}^0(b, p_T)$,
i.e.
the ratio in the absence of final state interaction. We have found that 
our final result 
depends little on the form of $R_{AA}^0$ when we take a $p_T$-shift of the form 
$\delta p_T=p_T^{\alpha}/C$. We have also check that 
our results for $p_T > 5$ GeV/c are almost independent on the form of the shift.

\noi A reasonable description of the data \cite{19r} for all values of $p_T$ has been obtained in \cite{13r} using
\beq
\label{10e}
\left . \delta p_T = p_T^{3/2}\right / (20  \ {\rm GeV}^{1/2}) 
\eeq

\noi and the initial suppression $R_{AA}^0$ given by eq. (\ref{8e}). Note that, at low $p_T$, $\delta p_T$ has to 
decrease in order to match the dual parton model results. 
For obvious reasons $\delta p_T$ is expected to vanish at small $p_T$.\par

This $p_T$ shift can be used in the whole range of $p_T$
available. However, as mentioned above, one can also use a constant
$\delta p_T$ for $p_T \ \gsim\ 6 \div 7$~GeV, without any significant
change in the results \cite{13r}. Using (\ref{10e}) in the whole $p_T$ region and
a value of 
$\sigma = 1.3$~mb, 
we obtain the results shown in Figs.~1
and 2. 
In Fig.~1 we show our results for $R_{AA}(p_T, b)$, eq. (\ref{7e}), 
in the centrality bin 0-10 \%, using
$\sigma = 1.3$~mb. 
They agree quite well with the available data \cite{19r} 
in the whole $p_T$ region, including the soft one. The corresponding values of the
initial suppression $R_{AA}^0(p_T, b)$ are also shown.  
In Fig. 2 we present our results for the dependence of $R_{AA}$ on
the number of participants, $N_{part}$, for $p_T \geq 4$~GeV. 
The comparison with the data shows that our model reproduces
the centrality dependence.

\section{Extension of the model to the transverse\break\noindent motion}
\hspace*{\parindent} In its formulation, the final state interaction model introduced in
section 2 takes into account the longitudinal expansion -- with no
consideration for the motion in the transverse plane. Elliptic flow, on
the contrary, results from an asymmetry in the azimuthal angle, and,
thus, the motion in the  transverse plane plays a fundamental role. The
extension of the model to take this transverse motion into account is
by no means trivial. To the best of our knowledge a totally satisfactory
formulation is not available. In the following, we propose a simple extension of the model taking only into account the different
path length of the $\pi^0$ in the transverse plane for each value of
its azimuthal angle $\theta_R$ -- measured with respect to the reaction
plane. At $y^* \sim 0$, the path length $R_{\theta_R}$, measured from the center of the interaction region (overlap of the colliding nuclei) is given by
\beq
\label{11r}
R_{\theta_R}(b) = R_A {\sin (\theta_R - \alpha) \over \sin \theta_R}
\eeq

\noi where $R_A = 1.05 \ A^{1/3}$~fm is the nuclear radius and $\sin \alpha = b \sin  \theta_R/2R_A$. Note that eq. (\ref{11r}) is only valid in the region $0 \leq \theta_R \leq 90^{\circ}$, and the integral in $\theta_R$ from 0 to 360$^{\circ}$ is obtained by integrating from 0 to 90$^{\circ}$ and multiplying the result by four.

Our ansatz consists in the following replacement in eq. (\ref{2e})  
\beq
\label{12r}
\rho (b, y^*\sim 0) \to \rho (b, y^*\sim 0) R_{\theta_R}/<R_{\theta_R}>
\eeq

\noi where $< R_{\theta_R}> = \int_0^{90^{\circ}} d\theta_R\
R_{\theta_R}/\int_0^{90^{\circ}} d \theta_R$. This is motivated by the
fact that, the duration of the interaction, characterized by the
argument of the logarithm in eq. (\ref{2e}), as well as the density of
the medium traversed by the $\pi^0$, are expected to be proportional to
the $\pi^0$ path length associated to its azimuthal angle $\theta_R$,
inside the overlap region of the colliding nuclei. Due to the division by $<R_{\theta_R}>$ in (\ref{12r}), the results in Section 3 are practically unchanged.\par

With the replacement (\ref{12r}), the survival probability of the
$\pi^0$, $S_{\pi^0}^{\theta_R} (p_T,b)$, depends on the angle $\theta_R$
and the elliptic flow can be obtained as
\beq
\label{13r}
v_{2}(p_T, b) = {\int_0^{90^{\circ}} d \theta_R \ \cos 2 \theta_R\ S_{\pi^0}^{\theta_R} (p_T, b) \over \int_0^{90^{\circ}} d\theta_R\ S_{\pi^0}^{\theta_R}(p_T, b)}\ .
\eeq

\noi Clearly, when the $\pi^0$ moves along the reaction plane its path
length will have its minimal value and the survival probability its
maximal one. On the contrary, for $\theta_R = 90^{\circ}$ the path
length will be maximal and the survival probability minimal. Thus, the
resulting value of $v_{2}$ is positive.\par

Using the same value $\sigma = 1.3$~mb of the final state interaction
cross-section and the same $p_T$-shift introduced in section 3 in order to describe
the experimental values of $R_{AA}(b, p_T)$, we obtain the values of
$v_{2}$ versus $p_T$ and centrality shown in Figs.~3, 4 and 5.
These
results are compared with 
PHENIX and STAR data \cite{20r, 20rnew, 21r} 
for charged hadrons.
Notice that the results for identified pions are practically the same
at least in the low $p_T$ range
\cite{5r}.

It is quite satisfactory that our results are in agreement with experiment for central and medium central collisions, which contain the bulk of the data. The situation for peripheral collisions is further discussed in Section 5. We also describe $v_2$ in the low and intermediate $p_T$ region. At $p_T > 3$~GeV our results (dashed line) overestimate the data. A natural way to overcome this problem is discussed in the next section.

\section{The large $p_T$ region}
\hspace*{\parindent} 

Using eq. (\ref{8e}), we have obtained  an initial
suppression $R_{AA}^0(p_T)$
that increases rapidly towards unit with increasing $p_T$, reaches values substantially
larger than unity and does not decrease towards one at large $p_T$ (see Fig.~1). Such a behaviour is expected in the small $p_T$ region 
where $R_{AA}^0$ increases rapidly. As mentioned above
 part of this increase is due to the decrease of shadowing 
when $p_T$ increases and part of it to the Cronin effect.
 At large $p_T$, on the contrary, this behaviour is not the conventional one. 
One rather expects that at large $p_T$,
 say $p_T > 5$~GeV, where the shadowing corrections 
are already quite small, the value of $R_{AA}^0$ is close or equal
to unity -- consistent with the smallness of the Cronin effect 
observed in $d$-$Au$ as well as in the direct photon
production data in $Au$ $Au$.\par 

The value of the $\pi^0$ suppression, characterized by the ratio
(\ref{7e}) is computed by multiplying $R_{AA}^0(b, p_T)$  
by the $\pi^0$ survival probability $S_{\pi^0}$ due to final state
interaction, given by eq. (\ref{2e}). 
In view of that, it is not possible to determine $R_{AA}^0$ from the large $p_T$
 suppression data alone 
-- since one can change the two factors $R_{AA}^0$ and $S_{\pi^0}$, keeping their product 
unchanged. 
On the contrary, we have seen that the values of the elliptic flow $v_2$ depend only 
on $S_{\pi^0}$. 
Thus, the combined analysis of both sets of data allows a separate determination of the two factors. We are going to show that only values of $R_{AA}^0$ close to unity at large $p_T$ can describe both sets of data. Accordingly, we proceed as follows: we use eq. (\ref{8e}) only for small $p_T$ and
take $R_{AA}^0 (b,p_T \geq 5$~GeV) = 1. Since the large $p_T$ suppression is given by the product $R_{AA}^0(p_T)
 S_{\pi^0} (p_T)$,  this change of $R_{AA}^0$ for $p_T\geq 5$~GeV has to go along with a change in
$S_{\pi^0}(p_T)$ -- which can be achieved by changing the $p_T$-shift at large $p_T$. Let us 
parametrize it in the form $\delta p_T = p_T^{\alpha}/C$ with the parameters $\alpha$ 
and $C$ determined in
such a way that the size and (practically flat) shape of the large $p_T$ suppression data is preserved.
This is achieved with $\alpha = 0.8$ 
and $C = 9.5$ GeV$^{-0.2}$. Thus, we shall take
\beq
\label{11enew}
\delta p_T = p_T^{0.8}/9.5 \ {\rm GeV}^{-0.2} \quad \hbox{for $p_T \geq 5$ GeV}\ .
\eeq

Let us note that eqs. (\ref{10e}) and (\ref{11enew})  lead to the same value of the $p_T$-shift at
$p_T = 2.9$~GeV. Therefore, in order to have a smooth $p_T$-shift in the whole $p_T$ region, 
we are going
to use eq. (\ref{10e}) for $p_T < 2.9$~GeV and eq. (\ref{11enew}) for $p_T \geq 2.9$~GeV. 
In the first
region ($p_T < 2.9$~GeV) we use the form of the initial suppression $R_{AA}^0(p_T)$ given by eq. (\ref{8e})
and, at large $p_T$, we take $R_{AA}^0 (p_T \geq 5$~GeV) = 1. 
In this way our results for $R_{AA}$ are practically the same,
as can been seen in Fig.~1, where we show the form of the two different parametrizations 
for the ratio $R_{AA}^0$ -- ratio before the interaction -- and the resulting $R_{AA}$.
Both parametrizations give similar results concerning the ratio for large $p_T$, for this reason
Fig.~2 remains practically unchanged.
However, the values of $v_2$ at large $p_T$ are substantially reduced. 
One obtains the full line in Fig.~3. 
This shows a saturation of $v_2$ at $p_T\ \gsim\ 3$~GeV which is consistent with experiment.\par

More information on the large $p_T$ values of $v_2$ and their centrality dependence is provided by the preliminary data on the azimuthal angle dependence $S_{\theta_R}(p_T)$ of the $\pi^0$ suppression at different centralities \cite{22r}. Our results are given in Fig.~6 
and compared with experiment. The agreement with the data is reasonable, indicating that the centrality dependence is approximately reproduced. This is in sharp contrast with the disagreement observed at lower values of $p_T$ (Fig.~4), where our centrality dependence is too weak as compared with the data
but in much
better agreement with the ones in Fig. 5. This important point has to be clarified.\par

In the above calculation of $v_2$ we have evaluated the path length of the $\pi^0$ (eq. (\ref{11r}) at its most probable position in the transverse plane, namely the center of the interaction region determined by the overlap of the two colliding nuclei. In a more accurate calculation, we should evaluate this path length at the point in the transverse plane where the $\pi^0$ has been produced. However, we have estimated that the error induced by this approximation is less than 10 \%. Indeed, we have checked numerically that moving the $\pi^0$ production point along the $\theta_R = 0$ direction the value of $v_2$ decreases, while, moving it in the $\theta_R = 90^{\circ}$ one, it increases. The average of these two values of $v_2$ is close to the one obtained using (\ref{11r}) -- although somewhat smaller. The difference increases when one moves away from the center and reaches the 10 \% level at the mid points on the two axis. Due to the integration over $T_{AA}(b,s)$ present in our calculation, the average position of the produced $\pi^0$ is closer to the center of the interaction region than to the mid-points. Thus, the estimated error is less than 10 \%.\par

While the approximation discussed above overestimates slightly the value of $v_2$, there is a physical effect which can increase its value in a more substantial way. Namely, if, besides the $p_T$-shift, the final state interaction produces also a shift $\delta\theta_R$ in the azimuthal angle $\theta_R$ of the $\pi^0$, it can be shown that the value of $v_2$ increases, irrespective of the sign of $\delta \theta_R$. For instance with $|\delta \theta_R| = 15^{\circ}$, the increase of $v_2$ at $b = 6$~fm and $p_T = 1.35$~GeV can be as large as 30 \%.\footnote{The effect depends on the time at which the shift $\delta\theta_R$ takes place. It is maximal when this time is close to freeze-out.}

\section{Conclusions}
\hspace*{\parindent} We have proposed a final state interaction model which takes into
account the different path length of a particle in the transverse
plane for each value of its azimuthal angle in the overlap area of the
colliding nuclei. The model is an extension of one previously
introduced to describe stangeness enhancement, $J/\psi$ suppression and
fixed $p_T$ suppression. Using the values of the final state
interaction cross-section (about 1~mb) and of the $p_T$-shift which describe the RHIC data on fixed $p_T$ $\pi^0$ suppression,
we obtain values of the elliptic flow $v_2$ close
to the measured ones for all values of $p_T$, including the soft
region. As for the centrality dependence, the values of $v_2$ are well
reproduced in the region $b\ \lsim\ 6$~fm which contains the bulk of
the data. We argue in Section 5 that the situation for peripheral collisions has to be clarified.\par

Moreover, combining the analysis of large $p_T$ suppression and elliptic flow data,
it is possible to obtain information on
 the initial suppression, $R_{AA}^0(p_T)$, i.e.
the one in the absence of final state interaction -- which results
from
 the combined effect of shadowing and the Cronin effect. Values of $R_{AA}^0$ close to unity at large $p_T$ ($p_T \ \gsim \ 5$~GeV) are required in order to describe both data sets.\par

The mass dependence 
of $v_2$ is reproduced quite well. The decrease
of $v_2$ with increasing mass is just a consequence of the broadening 
of the $p_T$ distributions when the mass increases. It will be described in detail in a forthcoming
publication \cite{Ele}.

In view of the uncertainties discussed at the end of section 5 we have 
restricted our calculation to the second order coefficient $v_2$ and 
to
mid-rapidities. However, our formalism allows to understand
qualitatively the strong decrease of $v_2$ when moving away from
mid-rapidities. This decrease follows the decrease of
the medium density 
as expected in our approach.
Thus, we expect a dramatic
decrease of $v_2$ when moving away from mid-rapidities.\par

We do not claim that the mechanism we have introduced gives the only contribution to the elliptic flow. The latter is a
very subtle observable to which presumably different mechanisms do
contribute. In our opinion, our knowledge of the dynamics of the nuclear
interaction is not sufficient to disentangle all these mechanisms and,
therefore, to allow to draw clear-cut conclusions regarding the
interpretation of the measured values of $v_2$. \par

Together with previous work on strangeness enhancement, $J/\psi$
suppression and fixed $p_T$ suppression, all of which can be described
in our framework with final state interaction cross-section smaller than or
of the order of 1~mb, this work lends support to the idea that, despite
the large densities reached in central $Au$ $Au$ collisions, the final
state interaction is rather weak. In terms of string models, it means that there is ``cross-talk''
between different strings. However, the concept of string remains
useful and these models can be used to compute the densities needed as
initial conditions in the gain and loss differential equations which
govern the final state interaction. Such a weak cross-talk, in the
presence of many strings per unit of transverse are, is supported
theoretically by the small transverse size of the string -- with a 
radius of the order of 0.1~fm. On the experimental side it is
supported by the nuclear transparency of proton-nucleus collisions --
where no ``cross-talk'' is needed to reproduce the data. As a consequence, only rear events are substantially affected but the
bulk of the system is not -- indicating that the strength and duration time of the final state interaction are not
enough to drive the system to thermal equilibration.\\

\noi {\Large \bf Acknowledgments} \par\nobreak
It is a pleasure to thank N. Armesto, C. Pajares and C. Salgado for interesting discussions and
helpful suggestions. We also thank A. Krzywicki and D. Schiff for discussions. E. G. Ferreiro thanks the
Service de Physique Th\'eorique, CEA, Saclay, for
hospitality
during the completion of this work.

\begin{figure}
\centering\leavevmode
\ \ \ \
\vskip 1.5cm
\epsfxsize=6in\epsfysize=6in\epsffile{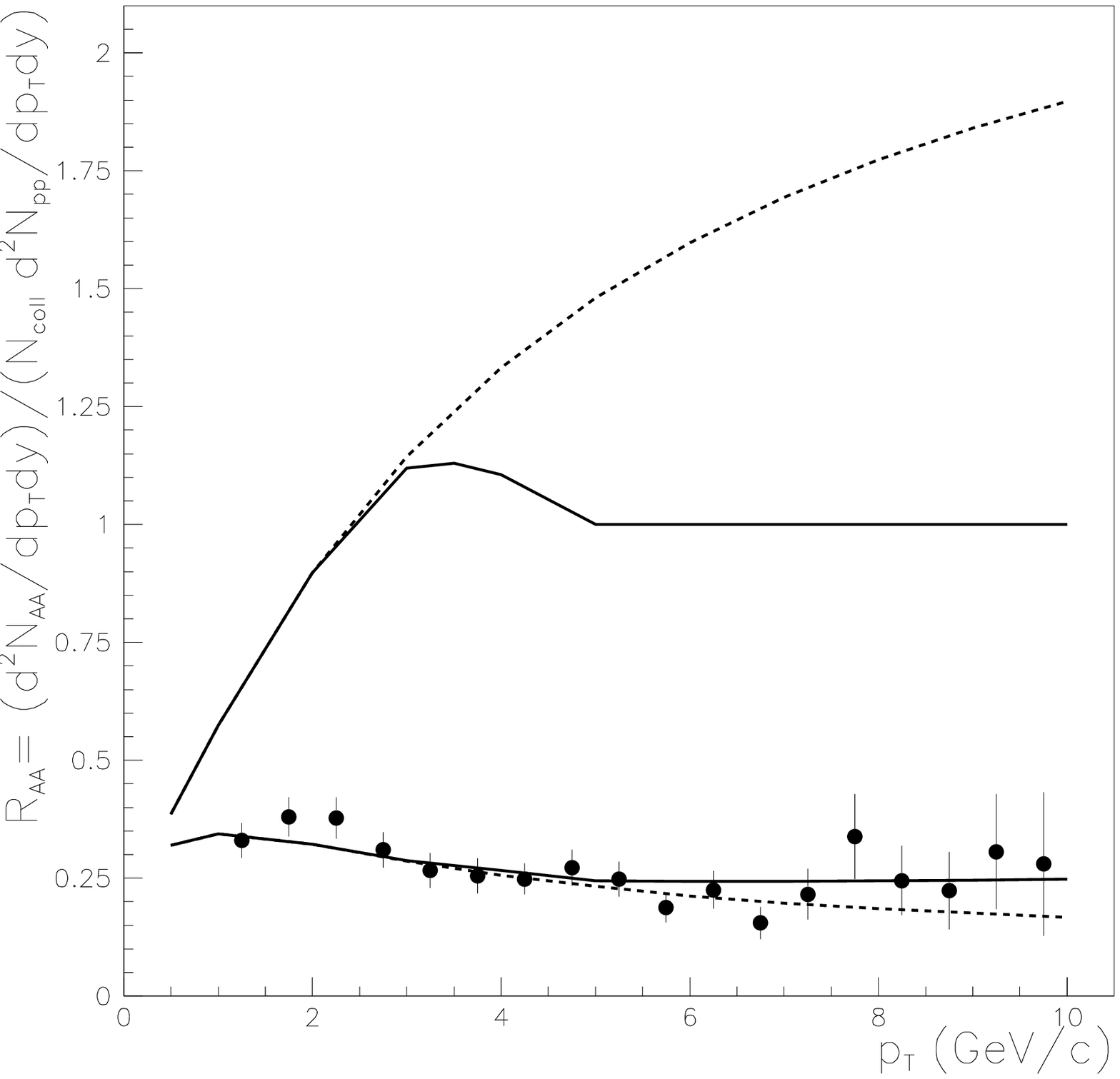}
\caption{The $\pi^0$ suppression factor $R_{AuAu}(p_T,b)$ at $\sqrt{s} = 200$~GeV, eq.
(\ref{7e}),
versus $p_T$ in the centrality bin 0-10 \% and the corresponding $\pi^0$ initial suppression factor
 $R_{AuAu}^0(p_T,b)$
(i.e. in the absence of final state interaction).
The dashed lines correponds to our results obtained using eq. ({\protect\ref{8e}}) for $R_{AuAu}^0$
and a $p_T$-shift given by eq. ({\protect\ref{10e}}). 
The continous lines are obtain with $R_{AuAu}^0$ described in Section 5 and the $p_T$-shift given 
eq. ({\protect\ref{10e}}) for $p_T < 2.9$~GeV and eq. ({\protect\ref{11enew}}) for $p_T \geq 2.9$~GeV.
The data are from {\protect\cite{19r}}.}
\end{figure}
 
\newpage
 
\begin{figure}
\centering\leavevmode
\ \ \ \
\vskip 1.5cm
\epsfxsize=6in\epsfysize=6in\epsffile{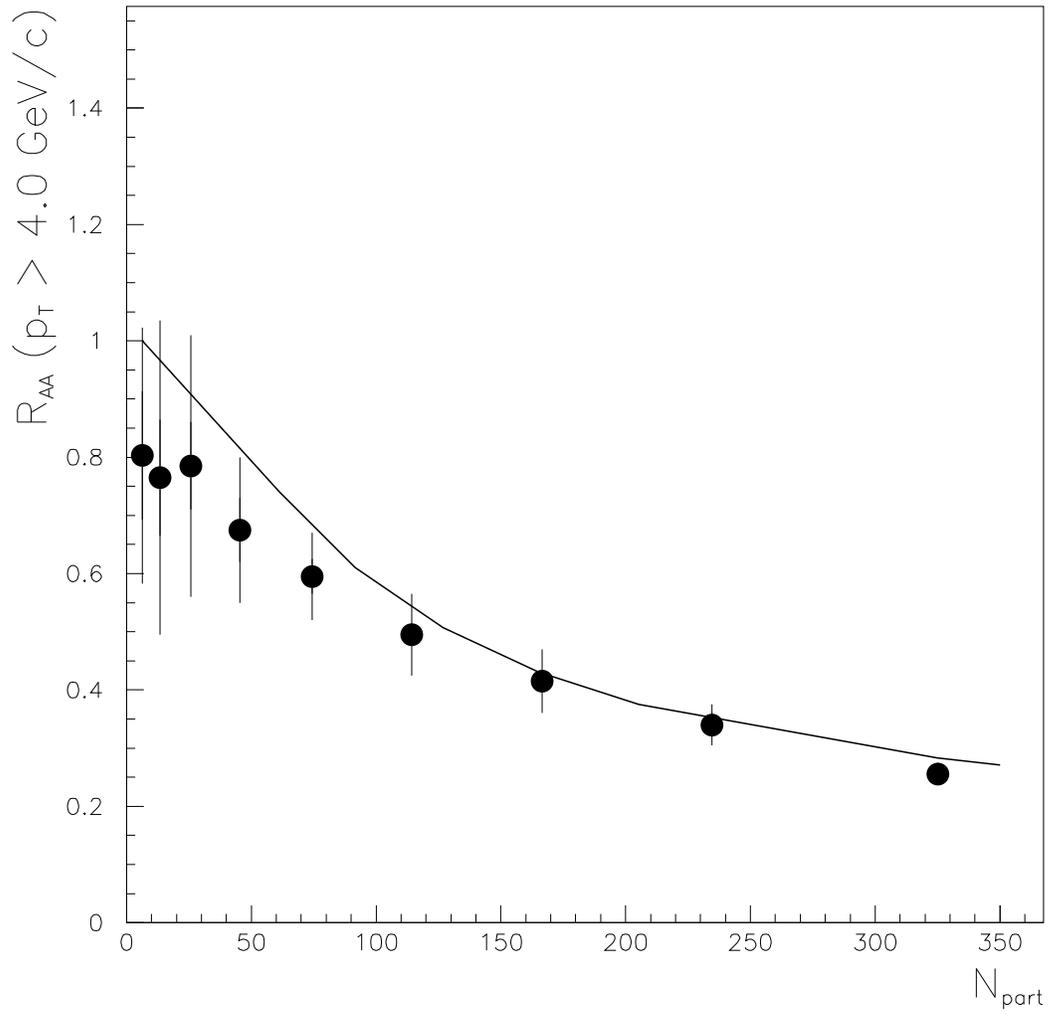}
\caption{
 The $\pi^0$ suppression factor $R_{AuAu}(p_T,b)$ versus the number of participants, 
$N_{part}$, for $p_T \geq 4$ GeV.
The data are from {\protect\cite{19r}}.}
\end{figure}

\newpage

\begin{figure}
\centering\leavevmode
\ \ \ \
\vskip 1.5cm
\epsfxsize=6in\epsfysize=6in\epsffile{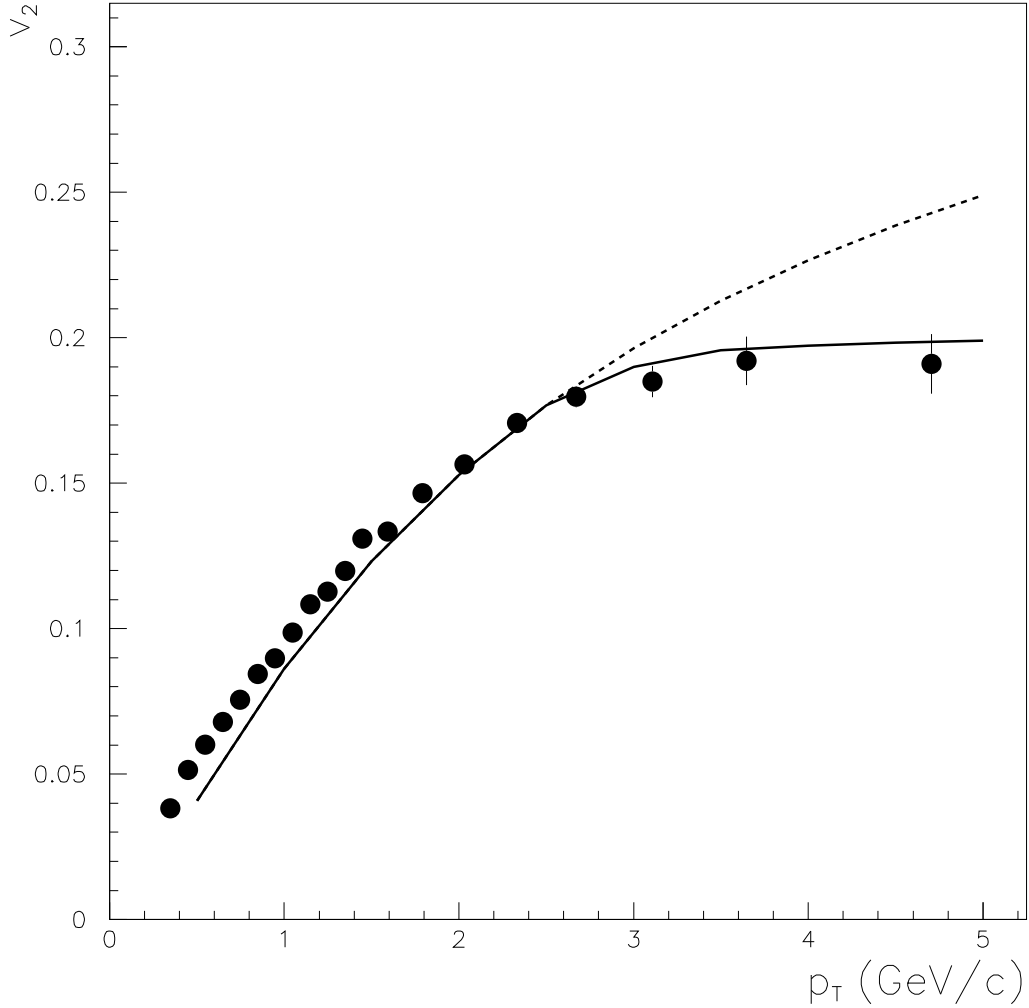}
\caption{
Values of the elliptic flow $v_2(b, p_T)$ versus $p_T$ in the centrality bin 13 \%-26~\%. 
The continous line correponds to our results obtained using eq. ({\protect\ref{11enew}}), 
the dashed line corresponds to our results
obtained using eq. ({\protect\ref{10e}}).
The data are from 
{\protect\cite{20r}}.}
\end{figure}

\newpage

\begin{figure}
\centering\leavevmode
\ \ \ \
\vskip 1.5cm
\epsfxsize=6in\epsfysize=6in\epsffile{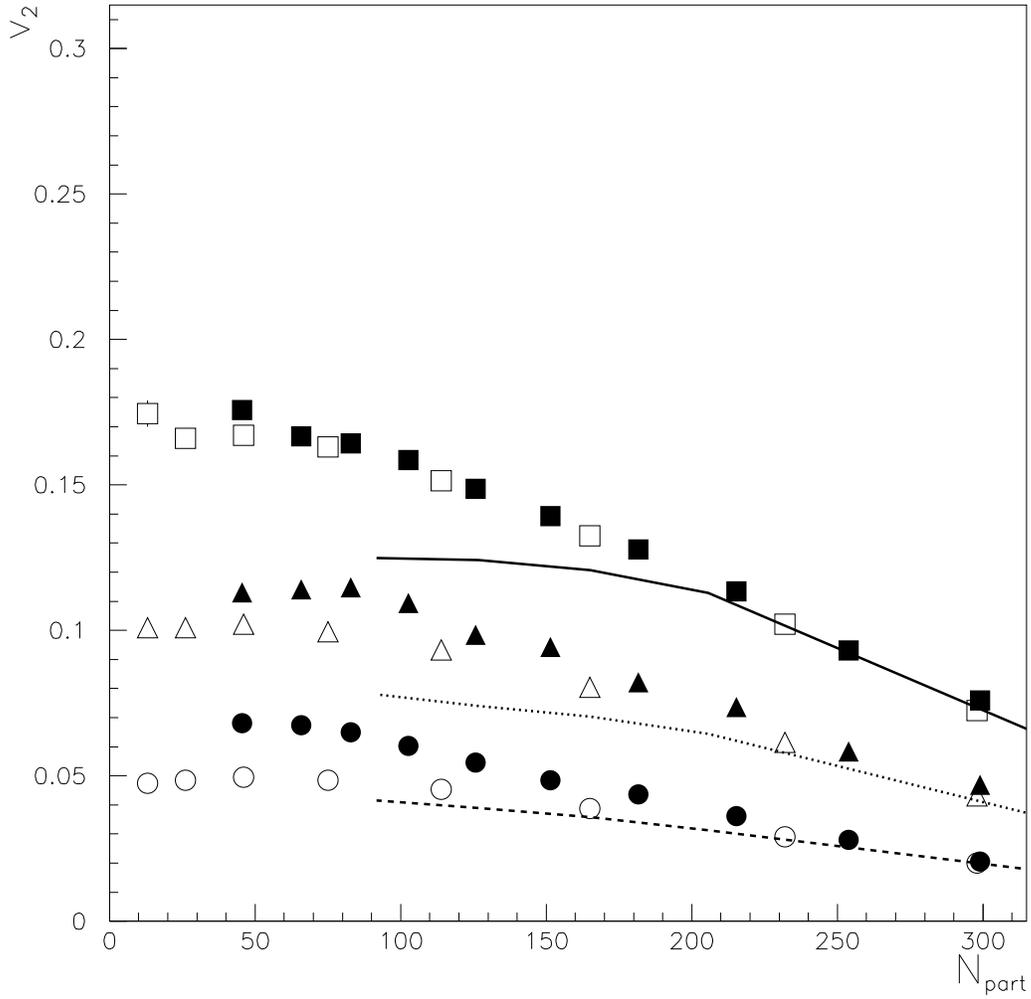}
\caption{
 Values of the $v_2(b, p_T)$ versus the number of participants for different values
of $p_T$~: $p_T = 0.4$~GeV (lower line), $p_T = 0.75$~Gev (middle) and $p_T = 1.35$~GeV (top). 
The black data are from PHENIX {\protect\cite{20r}}, the open data are from STAR 
{\protect\cite{21r}}.}
\end{figure}

\newpage

\begin{figure}
\centering\leavevmode
\ \ \ \
\vskip 1.5cm
\epsfxsize=6in\epsfysize=6in\epsffile{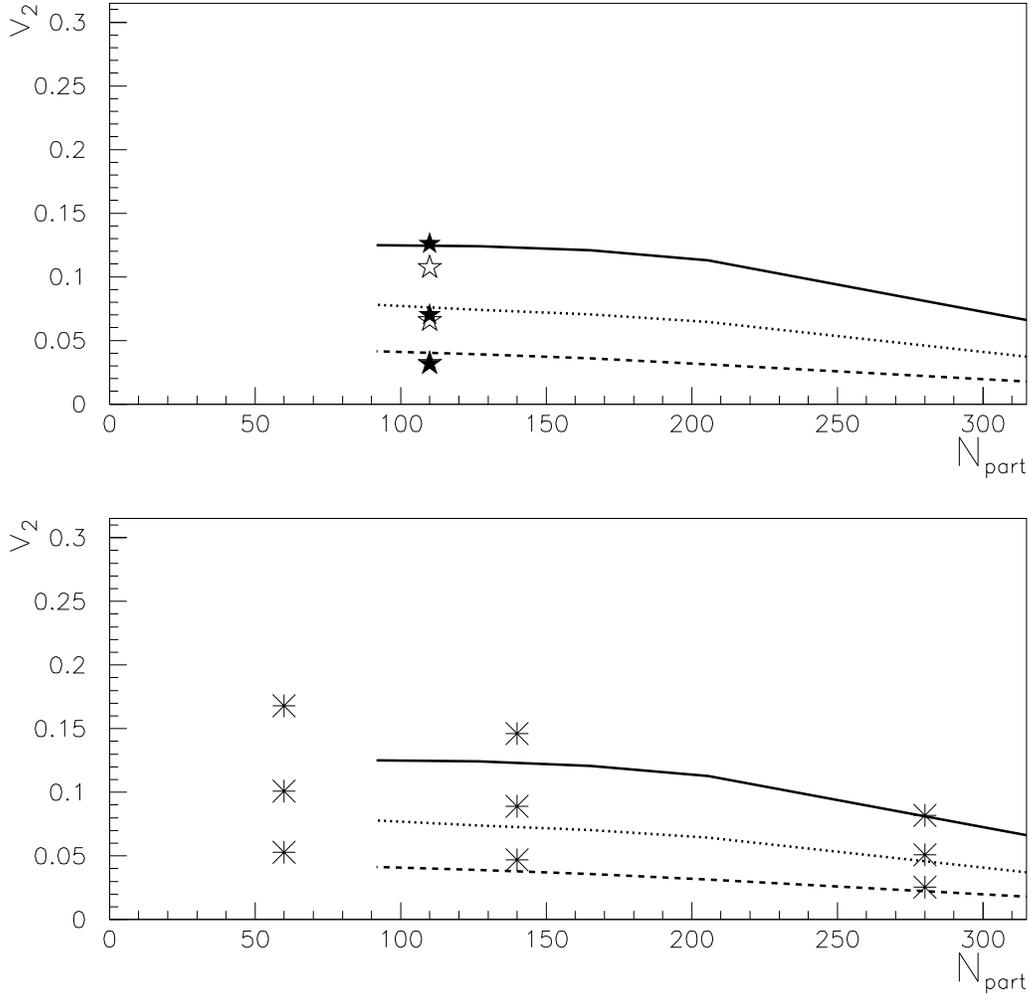}
\caption{
Values of the $v_2(b, p_T)$ versus the number of participants for different values
of $p_T$~: $p_T = 0.4$~GeV (lower line), $p_T = 0.75$~Gev (middle) and $p_T = 1.35$~GeV (top)
compared to:
data in minimum bias collisions from PHENIX {\protect\cite{20rnew}} 
(black symbols) and 
STAR {\protect\cite{21r}} (opened symbols) --{\it above}--, 
data from PHENIX {\protect\cite{20rnew}} --{\it below}--.}
\end{figure}

\newpage

\begin{figure}
\centering\leavevmode
\ \ \ \
\vskip 1.5cm
\epsfxsize=6in\epsfysize=6in\epsffile{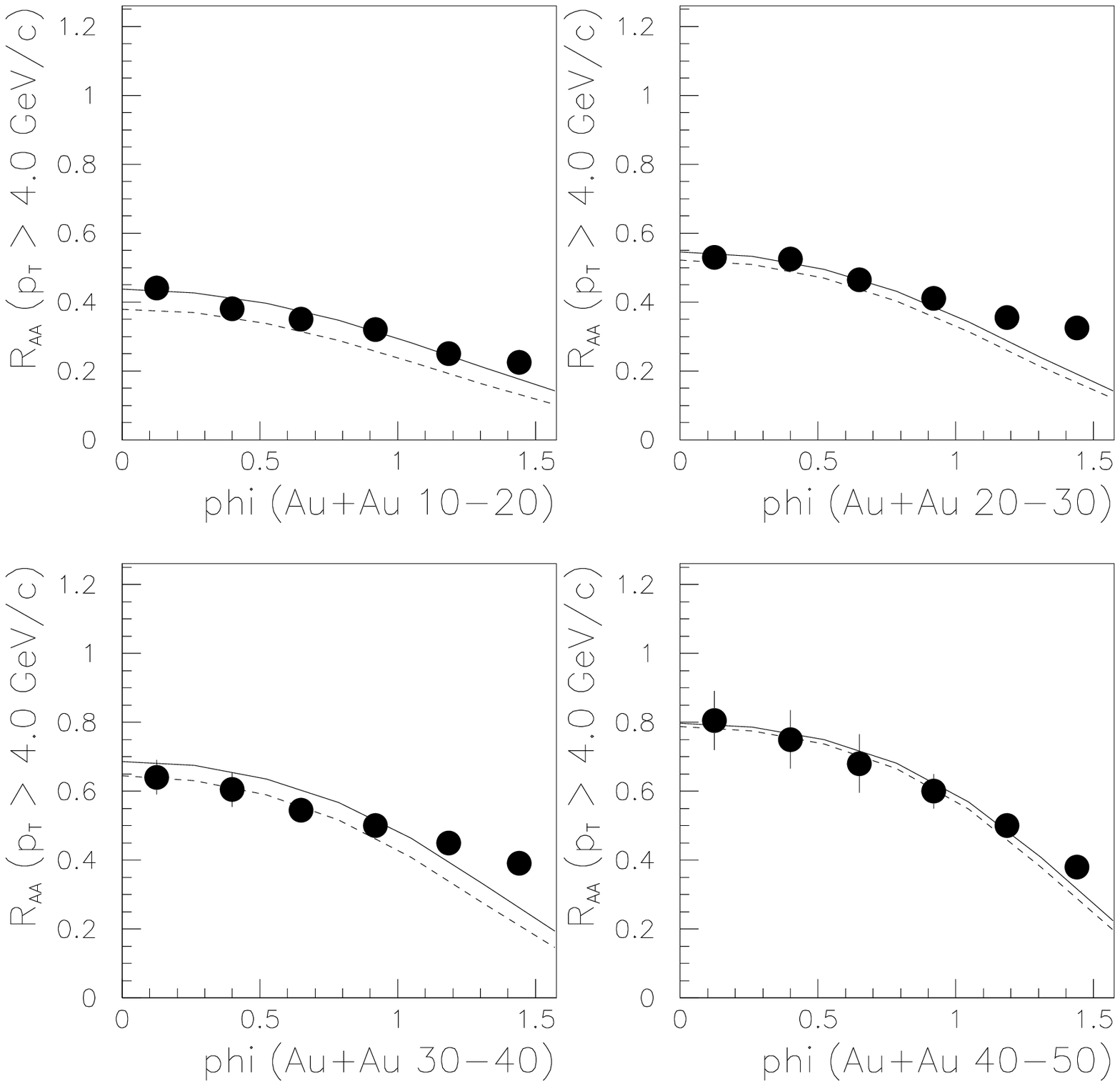}
\caption{
Values of the $\pi^0$ suppression $R_{AuAu}(b, \theta_R)$ as a function of the azimuthal angle $\theta_R$ (measured
from the impact parameter direction) for $p_T \geq 4$~GeV in various centrality bins. 
The continous line correponds to our results obtained using eq. ({\protect\ref{11enew}}),
the dashed line corresponds to our results
obtained using eq. ({\protect\ref{10e}}).
The preliminary data are from
{\protect\cite{22r}}.}
\end{figure}

\newpage


\newpage
\begin{thebibliography}{99}
\bibitem{1r} DPM : A. Capella, U. Sukhatme, C.-I. Tan, J. Tran Thanh Van, Phys. Lett. {\bf B81}, 68 (1979)~~; Phys. Rep. {\bf 236}, 225 (1994).\\
QGSM : A. Kaidalov, K. A. Ter-Martirosyan, Yad. Fiz. {\bf 39}, 1545 (1984)~; Yad Fiz. {\bf 41}, 1278 (1985).

\bibitem{2r} A. Tounsi, K. Redlich, hep-ph/0111159. \\
J. S. Hamich, K. Redlich, A. Tounsi, Phys. Lett. {\bf B486}, 61 (2000) and J. Phys. G {\bf 27}, 413 (2001).

\bibitem{3r} D. Kharzeev, M. Nardi, H. Satz, Phys. Lett. {\bf B405}, 14 (1997).\\
N. Armesto, A. Capella, E. G. Ferreiro, Phys. Rev. {\bf C59}, 395 (1999). \\
A. K. Chaudhuri, nucl-th/0212046.

\bibitem{4r} R. S. Bhalerao, J. P. Blaizot, N. Borghini and J. Y. Ollitrault, 
Phys. Lett. {\bf B627}, 49 (2005).

\bibitem{5r} STAR Collaboration, J. Adams et al., nucl-ex/0409033.

\bibitem{6r} P. F. Kolb, U. Heinz, nucl-th/0305084.

\bibitem{7r} D. Molnar, S. A. Voloshin, Phys. Rev. Lett. {\bf 91}, 092301 (2003).\\
V. Greco, C. M. Ko, P. Levai, Phys. Rev. Lett. {\bf 90}, 202302 (2003).\\
R. J. Fries, B. M\"uller, C. Nonaka, S. A. Bass, Phys. Rev. Lett. {\bf 90}, 202303 (2003).\\
R. C. Hwa, C. B. Yang, Phys. Rev. {\bf C70}, 024904 (2004).\\
D. Molnar, nucl-th/0408044. \\
S. Pratt, S. Pal, Nucl. Phys. {\bf A749}, 268 (2005); Phys. Rev. {\bf C71} 014905 (2005). 

\bibitem{8r} M. Bleicher, H. St\"ocker, Phys. Lett. {\bf B526}, 309 (2002).\\
Z. W. Lin, C. M. Ko, B. A. Li, B. Zhang, S. Pal, 
 Phys. Rev. {\bf C72}, 064901 (2005).\\
B. Zhang, M. Gyulassy, C. M. Ko, Phys. Lett. {\bf B455}, 45 (1999).

\bibitem{9r} A. Capella, C. A. Salgado, D. Sousa, Eur. Phys. J {\bf C30}, 111 (2003).

\bibitem{10r} A. Capella, D. Sousa, Eur. Phys. J. {\bf C30}, 117 (2003).\\
A. Capella, E. G. Ferreiro, A. Kaidalov, Phys. Rev. Lett. {\bf 85}, 2080 (2000).

\bibitem{11r} STAR Collaboration, C. Adler et al., Phys. Rev. Lett. {\bf 87}, 262302 (2001)~; J. Castillo, Nucl. Phys. {\bf A715}, 518c (2003).

\bibitem{12r} NA50 Collaboration, L. Ramello, Nucl. Phys. {\bf A715}, 243c (2003)~; L. Kluberg in Proc. Quark Matter 2004, Oklahoma, USA.

\bibitem{13r} A. Capella, E. G. Ferreiro, A. Kaidalov, D. Sousa, Eur. Phys. J {\bf C40}, 129 (2005).

\bibitem{14r} A. Drees, H. Feng, J. Jia, 
Phys. Rev. {\bf C71}, 034909 (2005).

\bibitem{15r} A. Dainese, C. Loizides, G. Pai\'e, 
Eur. Phys. J {\bf C38}, 461 (2005).

\bibitem{16r} B. Koch, U. Heinz, J. Pitsut, Phys. Lett. {\bf B243}, 149 (1990).

\bibitem{18r} A. Capella, E. G. Ferreiro, Eur. Phys. J {\bf C42}, 419 (2005).

\bibitem{17r} C. W. Jager, H. De Vries, C. De Vries, Atomic Data and Nuclear Tables {\bf 14}, 485 (1974).

\bibitem{19rnew} PHENIX Collaboration, S. S. Adler et al., Phys. Rev. {\bf C69}, 034910 (2004).

\bibitem{19r} PHENIX Collaboration, S. S. Adler et al., Phys. Rev. Lett. {\bf 91}, 072301 (2003).

\bibitem{20r} PHENIX Collaboration, S. S. Adler et al., 
Phys. Rev. Lett. {\bf 94}, 232302 (2005). 

\bibitem{20rnew} PHENIX Collaboration, S. S. Adler et al.,
Phys. Rev. Lett. {\bf 91}, 182301 (2003).

\bibitem{21r} STAR Collaboration, J. Adams et al., Phys. Rev. {\bf C72},
014904 (2005).

\bibitem{22r} PHENIX Collaboration, D. d'Enterria, Journ\'ee Th\'ematique 
on Jet quenching
IPN-Orsay, November 14th, 2005; Eur. Phys. J {\bf C43}, 295 (2005).

\bibitem{Ele} A. Capella and E. G. Ferreiro, Proceedings of XLIst 
Rencontres de Moriond: QCD and High Energy Hadronic Interactions, 
hep-ph/0604184.


 \end{thebibliography}
\end{document}